\documentclass[structabstract]{aa}  
\usepackage{graphicx}
\usepackage{txfonts}
\begin{document}
   \title{2D non-LTE radiative modelling of He I spectral lines formed in solar prominences}

   \author{L. L\'eger
          \and
          F. Paletou
          }

          \institute{Laboratoire d'Astrophysique de Toulouse-Tarbes, Universit\'e de Toulouse, CNRS, 14 av. E. Belin, 31400 Toulouse, France\\
            \email{[lleger;fpaletou]@ast.obs-mip.fr} }


 
  \abstract
  {The diagnosis of new high-resolution spectropolarimetric
    observations of solar prominences made in the visible and
    near-infrared mainly, requires a radiative modelling taking into
    account for both multi-dimensional geometry and complex atomic
    models.}
  {Hereafter we contribute to the improvement of the diagnosis based
    on the observation of He\,{\sc i} multiplets, by considering 2D
    non-LTE unpolarized radiation transfer, and taking also into
    account the atomic fine structure of helium.}
  {It is an improvement and a direct application of the multi-grid
    Gauss-Seidel/SOR iterative scheme in 2D cartesian geometry
    developed by us.}
  {It allows us to compute realistic emergent intensity profiles
    for the He\,{\sc i} $\lambda 10830$~\AA\, and $D_{3}$ multiplets,
    which can be directly compared to the simultaneous and
    high-resolution observations made at TH\'eMIS. A preliminary 2D
    multi-thread modelling is also discussed.}
   {}


   \maketitle
%

\section{Introduction}

Solar prominences (filaments) are made-up of dense and cool
chromospheric plasma hanging in the hot and low density corona
(Tandberg-Hanssen~\cite{tand}). Besides its intrinsic interest as a
natural laboratory for plasma physics, the study of these structures
is also of general interest in the frame of space weather
studies. Indeed, among other closed magnetic regions such as active
regions, eruptive prominences are often associated with coronal mass
ejections, or CMEs, that are huge plasma ``bubbles'' ejected from the
solar corona and able to strongly affect Sun-Earth relationships, by
their interactions with the terrestrial magnetosphere (see e.g.,
Gopalswamy et al.~\cite{gopalswamy} for a recent review upon the
various precursors of CMEs).

Despite systematic observations made since the nineteenth century and
decades of study, prominence formation mechanisms are still not well
understood. In particular, yet no theory can fully explain their
remarkable stability in a hotter and less dense medium. However, since
the plasma $\beta$ is low in prominences, the magnetic field is very
likely to play a major role in the physical scenarios which could
explain prominences formation, stability and, finally, the triggering
of these instabilities leading to CMEs.

He\,{\sc i} multiplets such as $\lambda 10830$~\AA\ in the
near-infrared, and the Fraunhofer ``yellow line'' $D_{3}$ at $\lambda
5876$~\AA~are the best tools, so far, to study prominence magnetic
fields. Indeed, only a few spectral lines are intense enough for
ground-based observations in the optical spectrum of solar prominences
i.e., at these wavelength at which spectropolarimetry is usually
done. These helium multiplets provide, even if they are fainter than
H$\alpha$ for instance, the most suitable information necessary for
the purpose of determining the magnetic field pervading the
prominence plasma, even though the helium spectrum, as observed at
high-spectral resolution, reveals atomic fine structure. First
spectropolarimetric observations of prominences and associated results
about the magnetic field properties have been reviewed by Paletou \&
Aulanier~(\cite{fpga}), for instance.

More recently, the first full-Stokes and high-spectral resolution
observations of the He\,{\sc i} $D_{3}$ multiplet made at TH\'eMIS
(Paletou et al.~\cite{fp01}) have led to a revision of magnetic field
inversion tools (L\'opez Ariste \& Casini~\cite{lopez}). In
particular, these authors demonstrated how taking into account
\emph{all} Stokes parameters, and not only linear polarization
signals, affects the reliability of the inversion
process. Furthermore, measurements of the ratio between the two peaks
resulting from the helium atomic fine structure of the He\,{\sc i}
$\lambda 10830$~\AA\, and $D_{3}$ multiplets (see Fig.11 in L\'opez
Ariste \& Casini~\cite{lopez} for instance) are often in contradiction
with the commonly used hypothesis of optically thin multiplets
(Bommier~\cite{bommier77}).

Besides, the most recent radiative models (Labrosse \&
Gouttebroze~\cite{lg01}, \cite{lg04}) still assume mono-dimensional
(1D) static slabs and \emph{no} atomic fine structure for the He\,{\sc i}
model-atom, which lead to the synthesis of unrealistic gaussian
profiles. It is therefore important to use the best numerical
radiative modelling tools in 2D geometry, as a first step, and a more
detailed He\,{\sc i} atomic model in order to improve our spectral
diagnosis capability. It is also a first application of the new 2D
radiative transfer code we have recently developed (Paletou \&
L\'eger~\cite{fp07}, L\'eger et al.~\cite{leger}).

We describe in Sect.~2 our prominence and our atomic model, as well as
our numerical strategy. Then, in Sect.~3 we compare and validate our
results against previous works of Labrosse \& Gouttebroze~(\cite{lg01},
\cite{lg04}). We also show how geometrical effects can influence line
profiles. In Sect.~4 we present 2D emergent intensity profiles for the
He\,{\sc i} $\lambda 10830$~\AA\, and $D_{3}$ multiplets which can be
compared to high-resolution spectropolarimetric observations made at
TH\'eMIS. Finally, we discuss some preliminary results of 2D
\emph{multi-thread} radiative models.


\section{2D non-LTE radiative modelling}

\subsection{Geometry and external illumination}

\begin{figure}
   \centering
   \includegraphics[width=6cm]{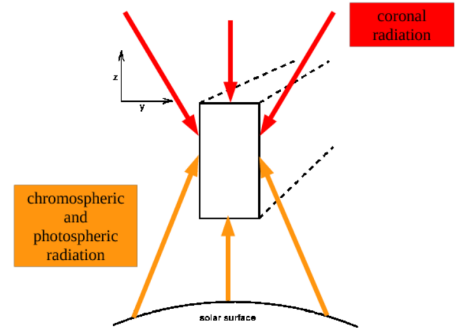}
   \caption{2D geometrical prominence model: the isolated, isothermal
     and isobaric slab is standing horizontally above the
     photosphere. For hydrogen and helium atoms, we consider a
     chromospheric and photospheric incident radiation coming from
     below, symmetrically on its sides and bottom. For the helium
     case, we consider in addition a constant coronal illumination
     coming from above onto the slab.}
         \label{model}
   \end{figure}

We have adopted the prominence geometrical model of
Vial~(\cite{vial}). As illustrated in Fig.~\ref{model}, it consists
in an isolated 2D slab standing above the photosphere. The
freestanding slab is supposed to be homogeneous, static, isothermal
and isobaric. It is characterized by two geometrical dimensions: its
horizontal thickness $D_{y}$ and its vertical extend $D_{z}$ (the
third dimension is infinite in 2D), and its altitude above the
photosphere $H_{0}$. The prominence material is then assumed to be
composed of neutral and ionized hydrogen, and helium.

The slab is illuminated from below, symmetrically on its sides and
bottom, by a photospheric and chromospheric incident radiation
field. The latter is diluted according to the chosen altitude above
the solar surface, as explained by Paletou~(\cite{fp96}). In addition,
for the helium case, we consider a \emph{constant} coronal
illumination on sides and top surfaces. Indeed, as shown by Andretta
\& Jones~(\cite{andretta}) and Mauas et al.~(\cite{mauas}), coronal
illumination plays a very important role in the formation of the
He\,{\sc i} multiplets, through the so-called
``photoionization--recombination'' (PR) mechanism. For the sake of
taking accurately into account for these external illumination
conditions, we have chosen depth points logarithmically spaced away
from the boundary surfaces towards the slab center, and symmetrically
distributed.

\subsection{Numerical strategy}

The numerical strategy used here is similar to the one described, in 1D, by Labrosse \& Gouttebroze~(\cite{lg01}, hereafter LG01):
   \begin{itemize}
   \item we solve the statistical equilibrium equations (SEE)
     including ionization balance, and the non-LTE radiative transfer
     equations self-consistently, for the multilevel hydrogen atom,
     using our 2D multi-grid GS/SOR iterative scheme (Paletou \&
     L\'eger~\cite{fp07}, L\'eger et al.~\cite{leger}). We thus obtain
     bound-level populations and electron densities (see also
     Paletou~\cite{fp95}, Heinzel~\cite{ph95} for the treatment of the
     ionization equilibrium). This sets physical conditions which
     shall be used as an input for the computation of the helium
     atom. The ratio between the helium and hydrogen total population
     densities is $\alpha=0.1$. We consider further that the helium
     atom is \emph{not} an electron donor in the prominence, so that
     the electron density given by the hydrogen computation is kept
     constant for the helium computation. Finally the Lyman continuum
     opacity and its emissivity are used to treat the helium
     ultra-violet (UV) lines pumped by the hydrogen continuum.
   \item Then, we solve self-consistently the SEE and the radiative
     transfer equations for the multilevel helium atom, using the same
     basic iterative scheme. To take into account the atomic fine
     structure, we consider the overlapping of all fine structure
     transitions into a single multi-component radiative transition. For
     example the $D_{3}$ multiplet is a combination of 5 fine
     structure transitions (see e.g., Fig.~3 in House \&
     Smartt~\cite{house}). We thus use, for the helium case, the
     Rybicki \& Hummer~(\cite{mali2}) \emph{full-preconditioning}
     strategy in order to solve the SEE including overlapping transitions.
   \end{itemize}

In our radiative transfer codes for the hydrogen and helium atoms,
we assumed complete redistribution in frequency (CRD).

\subsection{Atomic models}

We adopted the 5-level plus a continuum hydrogen atom model of
Paletou~(\cite{fp95}), whose data is consistent with earlier models of
Gouttebroze et al.~(\cite{ghv}).

For neutral helium, we considered two different cases: 19
bound-levels, up to $n=4$, plus a continuum, with \emph{no atomic fine
  structure} (hereafter our {\sc Hen4} model), or 17 bound-levels, up
to $n=3$, plus a continuum, with the atomic fine structure of
$2^{3}P$, $3^{3}P$ and $3^{3}D$ levels (hereafter, our {\sc Hen3sf}
model).

Energy levels and statistical weights were taken from the NIST
database (Ralchenko et al.~\cite{nist}). All other atomic data are
consistent with those of LG01: effective collision strengths,
collisional ionization coefficients and spontaneous emission
coefficients are from Benjamin et al.~(\cite{ben}), collision
strengths not defined there are from Benson \& Kulander~(\cite{bk}),
and photoionization cross sections are from TOPbase (Fernley et
al~\cite{fernley}).

Our radiative transfer code can explicitely consider collisional rates
among them and to/from all fine structure sublevels. For our {\sc
  Hen3sf} model, we set to 0 the rates between sublevels of the same
level. Otherwise, we used, for each sub-level, the rates given by the
later authors even though they were summed over the fine structure. We
are aware of this inconsistency which may have only a very small
influence on the results exposed hereafter.

Chromospheric illumination for all radiative transitions were taken
from Heasley et al.~(\cite{heasley}), with some additional details
which can be found in Labrosse et al.~(\cite{lg07}). Coronal
illumination for UV lines and continua were taken from
Tobiska~(\cite{tobi}) and Walhstr\o m \& Carlsson~(\cite{wc}).


\section{Preliminary results}

As a first step, results obtained with our 2D radiative transfer code
for the hydrogen atom are in very good agreement with those of Heasley
\& Milkey~(\cite{hm83}), Gouttebroze et al.~(\cite{ghv}) and
Paletou~(\cite{fp95})\footnote{In this article, partial redistribution
  in frequency (PRD) was used by default, but we could make direct
  comparisons with a CRD version of this antique code.}.

\begin{figure} 
\centering \includegraphics[width=8cm,height=11cm]{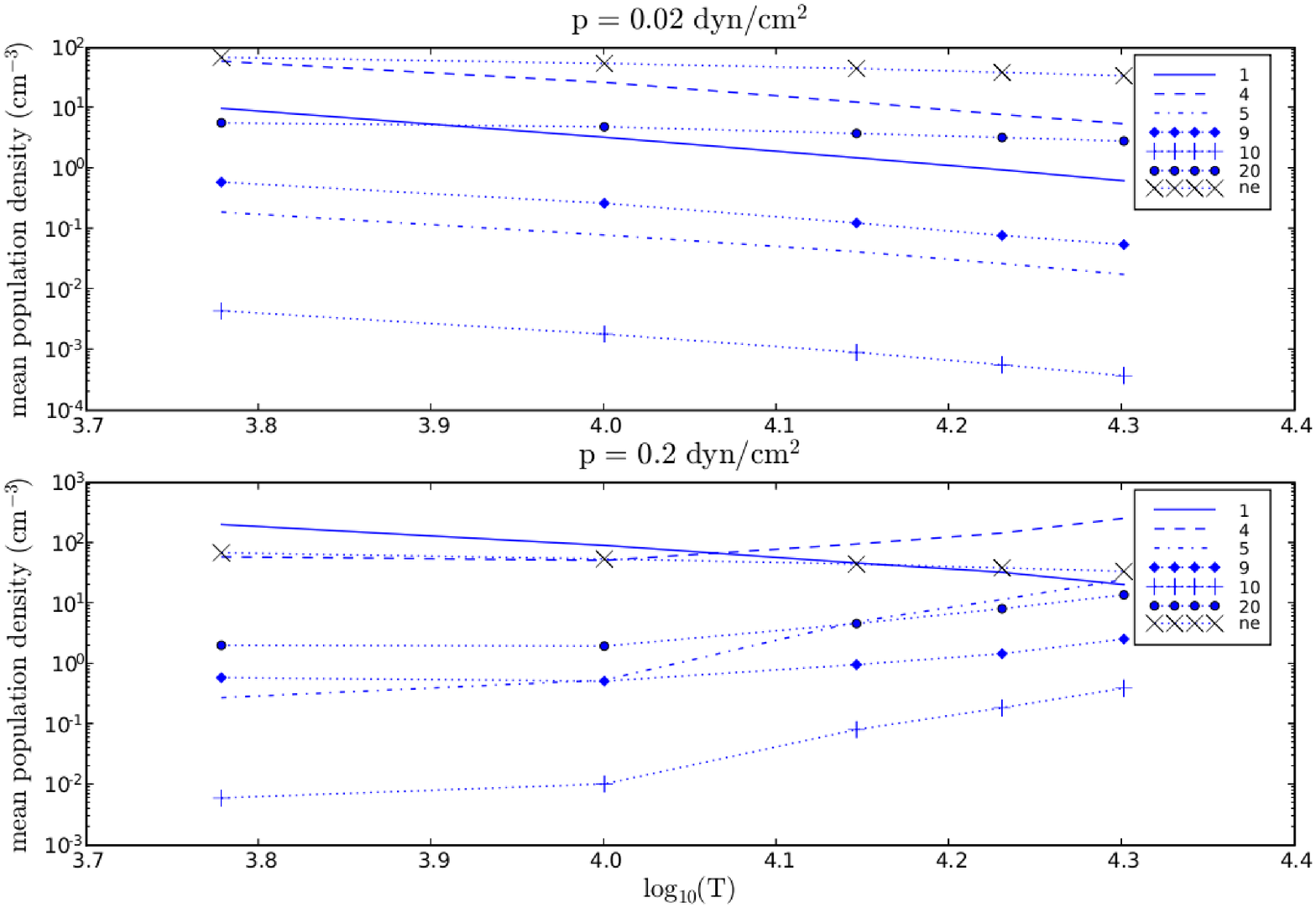}
\caption{Mean population and electron densities as a function
  of temperature for two pressures, $p_{g}=0.02$ (top) and $0.2$
  (bottom) $\mathrm{dyn}\, \mathrm{cm}^{-2}$. Numbers in the
  upper-right frames refer to the energy level as follows: 1 for the
  ground state of He\,{\sc i} and 20 the ground state of He\,{\sc
    ii}. \emph{The mean population densities for these two levels,
    as well as for the electron density, are divided by
    $10^{8}$}. The other codes for levels are: 4 for $2^{3}P$, 5 for
  $2^{1}P$, 9 for $3^{3}D$ and 10 for $3^{1}D$.}
\label{mean}
\end{figure}

As a second step, we have compared the results obtained with our
\emph{upgraded} 2D radiative transfer code for the helium atom with
those of LG01, as these last authors have made comprehensive
comparisons with the pionneering works of Heasley et
al.~(\cite{heasley}) and Heasley \& Milkey~(\cite{hm78}). In order to
do so, we have used our {\sc Hen4} model, without atomic fine
structure. Also, their radiative code is 1D plane parallel so that
they have neither coronal illumination on the top surface, nor
dilution effects varying with the altitude, as in 2D. However, their
helium atomic model is more detailed than ours, with 29 bound-levels
up to $n\,=\,5$ for He\,{\sc i}, 4 bound-levels for He\,{\sc ii}, and
He\,{\sc iii}. Unlike us, they also take into account PRD for the
resonance lines H\,{\sc i} Ly$\alpha$ and Ly$\beta$, He\,{\sc i}
$\lambda 584$~\AA\, and He\,{\sc ii} $\lambda 304$~\AA.

We have modeled a 2D prominence with $D_{z}\,=\,100\,000$ km and
$D_{y}\,=\,1\,000$ km. Its temperature, $T$, could take the respective
values: $6\,000$, $10\,000$, $14\,000$, $17\,000$ and $20\,000$ K, and
the gas pressure $p_{g}$ could take 2 values: $0.02$ and
$0.2\,\mathrm{dyn\,cm^{-2}}$. The microturbulent velocity was fixed at
$\xi = 5\,\mathrm{km}\,\mathrm{s}^{-1}$. The bottom of the slab was
set at $H_{0}=10\,000$ km above the solar surface. We have chosen a 2D
spatial grid with $123 \times 123$ logarithmically spaced points. We
have also used Doppler profiles monotonically sampled with a 0.1 step
in Doppler width units.

In Fig.~\ref{mean}, mean population densities variations against
temperature are displayed. Mean densities are defined here as:
$$N_{i}\,=\,\displaystyle{\frac{\int_{0}^{D_{y}}\,n_{i}(y,D_{z}/2)dy}{D_{y}}}\,,$$
where $n_{i}(y,D_{z}/2)$ is level $i$ population at the position $y$
at mid-height of the slab ($z=D_{z}/2$). We have focused on the
populations of 5 energy levels of He\,{\sc i}: $1s^{2}$, $2^{3}P$,
$2^{1}P$, $3^{3}D$, $3^{1}D$ labelled 1, 4, 5, 9 et 10, and the ground
state of He\,{\sc ii} labelled 20. For the sake of clarity, \emph{mean
  population densities of the ground states of He\,{\sc i} and
  He\,{\sc ii}, as well as the electron density, are divided
  by $10^{8}$ on the figure}.

\begin{figure}
\centering
\includegraphics[width=10cm]{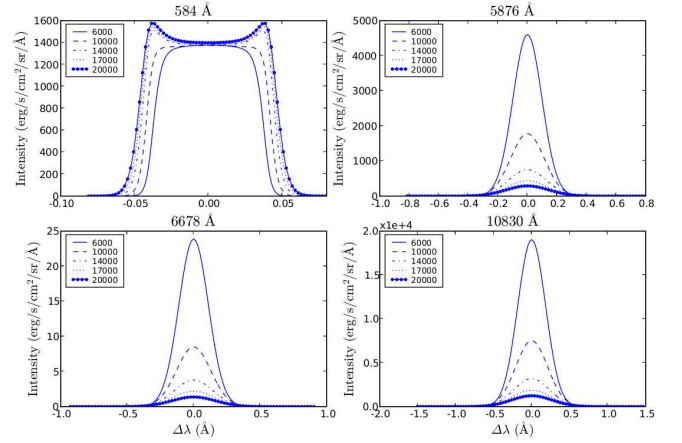}
\caption{Emergent intensities for lines He\,{\sc i} $\lambda 584$~\AA,
  $\lambda 5876$~\AA, $\lambda 6678$~\AA\, and $\lambda 10830$~\AA,
  for a pressure $p_{g}=0.02\,\mathrm{dyn}\, \mathrm{cm}^{-2}$,
  for 5 different temperatures, and for $D_{z}\,=\,100\,000$ km and
  $D_{y}\,=\,1\,000$ km. These intensities are computed at slab
  mid-height, $z=D_{z}/2$, for a line-of-sight perpendicular to the slab.}
\label{int1}
\end{figure}

\begin{figure}
\centering
\includegraphics[width=10cm]{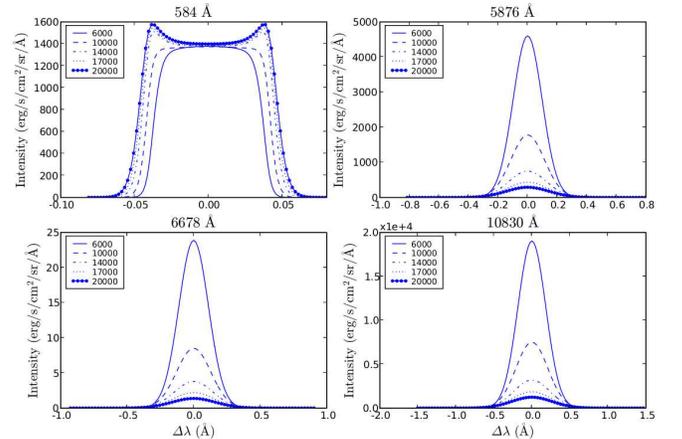}
\caption{Same as Fig.~\ref{int1} for a vertical geometrical extension
  $D_{z}\,=\,10\,000$ km.}
\label{int2}
\end{figure}

In the low pressure regime, $p_{g}\,=\,0.02\,\mathrm{dyn\,cm^{-2}}$,
we recover the same general behaviour as the one shown by in LG01: an
increase in temperature reduces the neutral helium mean population and
increases the ionized helium mean population, until the latter becomes
greater than the former. This transition temperature is around
$8\,000$ K for our model, whereas it is $11\,000$ K in LG01. This is
explained by PRD effects, as LG01 models in CRD give the same result
as ours (N. Labrosse, private communication).

For larger pressures, $p_{g}\,=\,0.2\,\mathrm{dyn\,cm^{-2}}$,
it is shown in Fig.~\ref{mean} that all mean population densities
increase with temperature but the ground state of He\,{\sc i}. As
emphasized indeed by LG01, the helium ionization is less important
than in the low pressure case, because of the optical thickness of the
helium continuum which prevents incident radiation from reaching the
core of the slab. We also recover this effect using our 2D code, when
the vertical extension of the slab is made large enough, mimicking a
1D vertical slab model though.

We could also check a posteriori, comparing n(He\,{\sc ii}) and
n$_{\rm e}$, that the assumption that He is not an electron donor
remains valid for most of the range of parameters we used. It becomes
however questionable for both high pressure and high temperature
cases, as shown in Fig.~\ref{mean}.

In Fig.~\ref{int1}, emergent intensities for lines He\,{\sc i}
$\lambda 584$~\AA, $\lambda 5876$~\AA, $\lambda 6678$~\AA\, and
$\lambda 10830$~\AA\, are displayed for a vertical extension
$D_{z}=100\,000$ km, for a pressure
$p_{g}=0.02\,\mathrm{dyn\,cm^{-2}}$ and for five different
temperatures. Intensities are computed at mid-height of the slab, at
$z=D_{z}/2$, and for a line-of-sight (los) perpendicular to it.

The general behaviour of these intensities with temperature is in
excellent agreement with results presented by LG01 (see e.g., their
Fig. 8). Our values for the intensity of the resonance line He\,{\sc
  i} $\lambda 584$~\AA\ are smaller than those of LG01, whereas they
are greater for the optically thin $\lambda 5876$~\AA\, and $\lambda
6678$~\AA\, lines. This could be explained by PRD effects which play
an important role in determining the shape of the resonance line
He\,{\sc i} $\lambda 584$~\AA\, as well as in the He\,{\sc i} levels 1
and 4 populations. These PRD effects have consequently an impact on
the line He\,{\sc i} $\lambda 5876$~\AA\, as its lower level is the
level 4, and on ionization via the neutral helium continuum $\lambda
504$~\AA. The ground state of He\,{\sc ii} mean population density is
smaller in LG01.  As the triplet levels are populated through the PR
mechanism (see Andretta \& Jones~\cite{andretta} and references
therein) from the ground state of He\,{\sc ii}, our $D_{3}$ multiplet
emergent intensity is greater than the ones of LG01.

In Fig.~\ref{int2}, the same emergent intensities are displayed for a
smaller vertical extension of $D_{z}\,=\,10\,000$ km (note also that
we used, in both case, the same fixed dilution factor for the incident
radiation, in order to put in evidence the only effects due to
different vertical slab extensions). Emergent intensity values are
larger, with relative differences ranging between 15 and 40\%, than
those displayed in Fig.~\ref{int1}. This shows that significant 2D
geometrical effects take place when one reduces the geometrical
extension of the slab, and how they impact the magnitude of emergent
intensities. Such kind of effects were first put in evidence on
H$\alpha$ by Paletou~(\cite{fp97}).


\section{A realistic spectral synthesis of He\,{\sc i} multiplets}

\begin{figure} \includegraphics[width=8cm,height=6.5cm]{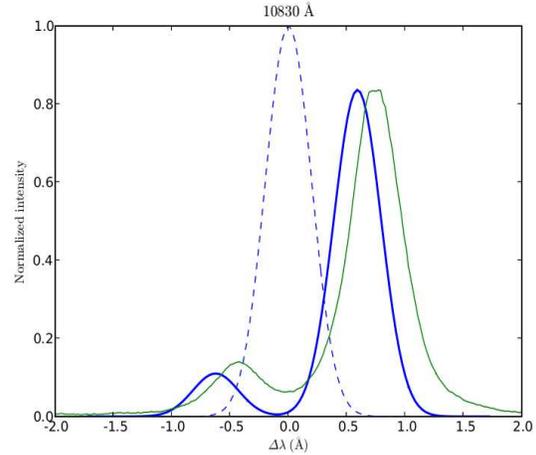}
  \caption{Normalized emergent intensities for the He\,{\sc i}
    $\lambda 10830$~\AA\, multiplet obtained with the {\sc Hen4}
    atomic model (dashed lines), with the {\sc Hen3sf} atomic model
    (thick line), and from TH\'eMIS observations of June 2007, vs. the
    wavelength shift with respect to the multiplet central (vacuum)
    wavelength defined as $\lambda=10832.7$~\AA. Synthetic profiles
    are for a line-of-sight perpendicular to the slab. Both the
    fine-structured and the observed profiles were normalized to the
    maximum amplitude of the {\sc Hen4} synthetic profile. The
    observed one was also slightly shifted in frequency, for the sake
    of comparison with the {\sc Hen3sf} one.}
\label{int10830}
\end{figure}

\begin{figure}
\includegraphics[width=8cm,height=6.5cm]{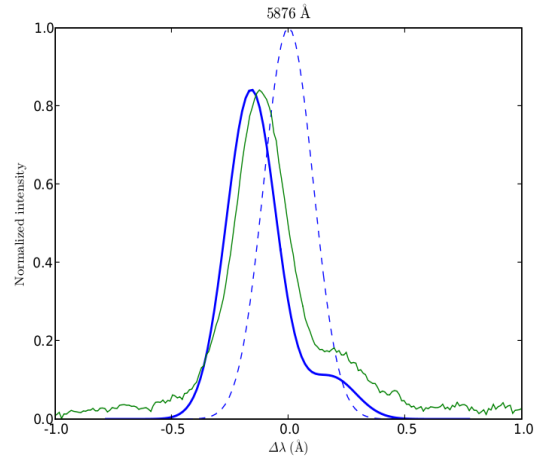}
\caption{Same as Fig.~\ref{int10830} for the He\,{\sc i} $D_{3}$
  multiplet whose central (vacuum) wavelength is $5877.4$~\AA.}
\label{intd3}
\end{figure}

We now focus on He\,{\sc i} $D_{3}$ and $\lambda 10830$~\AA\,
multiplets, as our primary aim is to model realistic emergent line
profiles consistent with high-resolution spectropolarimetric
observations made at the solar telescope TH\'eMIS (see e.g., Paletou
et al.~\cite{fp01}). With such an instrument, the two ``red'' and
``blue'' peaks are, indeed, clearly resolved (see also e.g., Merenda
et al.~\cite{laura}, Trujillo Bueno et al.~\cite{jtb02}, for data
taken at the German VTT with the TIP polarimeter, in the
near-infrared).

Hereafter, we shall use in our 2D radiative transfer code the {\sc
  Hen3sf} atomic model which takes into account the atomic fine
structure for the $2^{3}P$, $3^{3}P$ and $3^{3}D$ levels of He\,{\sc
  i}.

\subsection{Single slab models}

We have modeled a 2D isobaric, isothermic and static prominence with a
vertical extension $D_{z}=30\,000$ km and a horizontal extension
$D_{y}=5\,000$ km. Its temperature is $T=8\,000$ K, and the gas
pressure is $p_{g}=0.05\,\mathrm{dyn\,cm^{-2}}$ which are rather
typical values for the modelling of quiescent prominences (see e.g.,
Gouttebroze et al.~\cite{ghv}). The microturbulent velocity is $\xi =
5\,\mathrm{km}\,\mathrm{s}^{-1}$. The bottom of the slab is set at
$H_{0}=10\,000$ km above the solar surface. We have chosen a 2D
spatial grid with 243 x 243 logarithmically spaced points.

In Figs.~\ref{int10830} and~\ref{intd3}, emergent intensities for
He\,{\sc i} $\lambda 10830$~\AA\, and $D_{3}$ multiplets are
displayed. They were computed at the slab mid-height, $z=D_{z}/2$, for
a los perpendicular to the slab, and the specific intensity was
normalized to the maximum value obtained with model {\sc Hen4}. Then,
maximum intensity values are
$21871\,\mathrm{erg\,s^{-1}\,cm^{-2}\,sr^{-1}\,}$\AA$^{-1}$ for the
$\lambda 10830$~\AA\, multiplet and
$6473\,\mathrm{erg\,s^{-1}\,cm^{-2}\,sr^{-1}\,}$\AA$^{-1}$ for the
$D_{3}$ multiplet. Taking into account the atomic fine structure, we
obtain indeed, 2D emergent line profiles which are \emph{directly
  comparable} to high-resolution spectroscopic observations,
characterized by two well-resolved subcomponents.

We define $I_{r}$ as the ratio between the intensity of the largest
subcomponent peak and the intensity of the smaller peak, that is,
$I_{blue}/I_{red}$ for the $D_{3}$ multiplet and $I_{red}/I_{blue}$
for the $\lambda 10830$~\AA\, multiplet.  We remind here that, when
the He\,{\sc i} $D_{3}$ multiplet becomes optically thick, the $I_{r}$
ratio is lower than 8 (House \& Smartt~\cite{house}, Landi
Degl'Innocenti~\cite{landi}).  Furthermore, statistics made on this
ratio give a mean value of 6 (L\'opez Ariste \& Casini~\cite{lopez}),
which is not consistent with the commonly used hypothesis that the
$D_{3}$ multiplet is optically thin.

We can evaluate the optical thickness of the two multiplets obtained
with our 2D model. At the slab mid-height, $z=D_{z}/2$, and for the
frequency corresponding to the maximum intensity value:

\begin{eqnarray*}
\tau(10830) & = & 5,1\,\times\,10^{-2}\\
\tau(D_{3}) & = & 4,4\,\times\,10^{-3}\,.
\end{eqnarray*}
The two multiplets are thus optically thin, and we also find that
$\tau(10830)\,\sim\,10\,\times\,\tau(D_{3})$, as pointed out by Andretta \&
Jones~(\cite{andretta}). In these two cases, $I_{r}$ ratios are indeed around
8.

Having performed several tests with temperatures varying from $6\,000$
to $20\,000$ K and for gas pressures ranging from $0.02$ to
$0.2\,\mathrm{dyn}\, \mathrm{cm}^{-2}$, we found that the two
multiplets are optically thicker for both high pressure and high
temperature atmospheres. However, even in these regimes, the optical
thicknesses of the $D_{3}$ multiplet never go beyond
$\tau\,\sim\,0.1$. These results are also in accordance with 1D models
of Labrosse \& Gouttebroze~(\cite{lg04}). For instance, these authors
found no model for which the $D_{3}$ multiplet is optically thick, and
$\tau(10830)>1$ for models where $T>20\,000$ K and
$p_{g}>0.64\,\mathrm{dyn}\, \mathrm{cm}^{-2}$ i.e., conditions which
are not representative of those expected in quiescent prominences.

\emph{Which kind of model could reproduce the observed ratio between
subcomponents of the 557.6 and 1083. nm multiplets of He\,{\sc i}?}

\begin{figure}
\centering
\includegraphics[width=8cm]{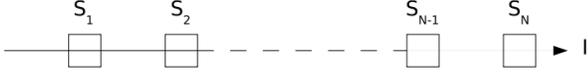}
\caption{A sketch of the 2D multi-thread prominence model we have
  adopted. Individual threads are superimposed along the line-of-sight,
  and there is no radiative interaction between them. Emergent
  intensities are computed as formal solutions of the radiative transfer
  equation along a line-of-sight perpendicular to the threads vertical
  extension, in order to take into account the contribution of each
  individual thread.}
\label{multi}
\end{figure}

\subsection{2D multi-thread models}

We have thus decided to model prominences differently, using the fact
that they are made of small-scale threads. This is supported both by
observations (e.g., Berger et al.~\cite{berger}, Lin et
al.~\cite{lin}) and by MHD simulations (e.g., Low \&
Petrie~\cite{low}). The effect of an increased penetration of the
incident radiation, using multiple layer modelling, was also shown in
1D by Gouttebroze et al. (\cite{goutte02}). Also, such a 2D non-LTE
\emph{spatial} fine-structure radiative modelling of solar prominences
was recently used by Gun\'ar et al. (2007), for the purpose of
interpreting UV observations of the Lyman series of hydrogen made with
the SUMER spectrograph on-board SoHO.

Hereafter, all the 2D threads (see Fig.~\ref{multi}) are supposed to
be identical and located at the same altitude, with similar horizontal
and vertical extensions $D_{y}=D_{z}=1\,200$ km which corresponds to a
$\approx 1.5$ arcsec angular resolution compatible with what can be
decently achieved while doing spectropolarimetry of prominences using
ground-based observatories. Such a choice can obvisouly be debated,
and we could have considered (much) smaller threads, beyond spatial
resolution (see e.g., Vial~\cite{vial06},
Heinzel~\cite{ph07}). However, we leave more detailed investigations
about such an issue to further studies.

We considered also that there is no radiative interaction between
individual threads. Thus, as a first step, we applied our 2D radiative
transfer code for one individual thread. Then we computed emergent
intensities as the formal solution of the radiative transfer equation
along a los perpendicular to each thread's vertical extension, in
order to take into account the contribution of \emph{a bunch of
  threads}. Therefore, the total optical thickness along the los is
simply the sum of the optical thicknesses of each individual thread.

\begin{figure} \includegraphics[width=8cm]{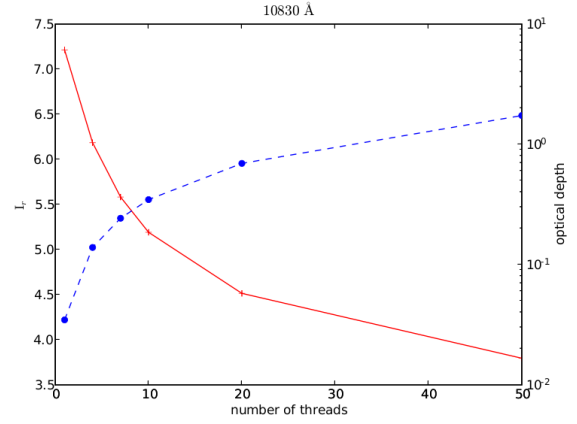}
  \caption{The $I_{r}$ ratio (solid line) and the optical thickness
    integrated on the slab height (dashed line), for the He\,{\sc i}
    $\lambda 10830$~\AA\, multiplet versus the number of individual
    threads, for a gas pressure $p_{g}\,=\,0.5\,\mathrm{dyn}\,
    \mathrm{cm}^{-2}$ and a temperature $T\,=\,8\,000$ K.}
\label{multi10830_8}
\end{figure}
\begin{figure}
  \includegraphics[width=8cm]{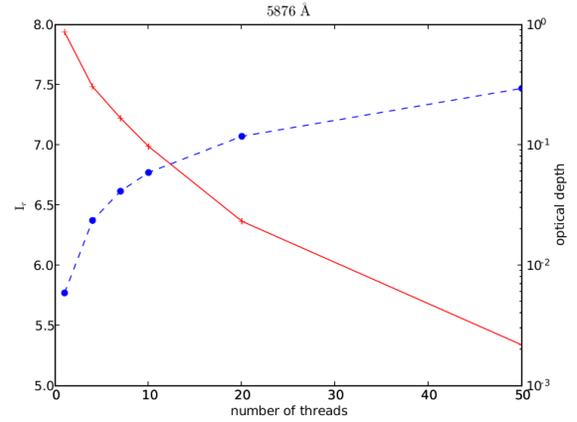}
  \caption{Same as Fig.~\ref{multi10830_8} for the $I_{r}$ ratio and
    the optical thickness of the He\,{\sc i} $D_{3}$ multiplet.}
\label{multi5876_8}
\end{figure}

For each (identical) thread, we adopted a gas pressure
$p=0.5\,\mathrm{dyn}\, \mathrm{cm}^{-2}$ and a temperature $T=8\,000$
K corresponding to a commonly used temperature for quiescent
prominences (see e.g., Engvold et al.~\cite{engvold},
Tandberg-Hanssen~\cite{tand}).

In Figs.~\ref{multi10830_8} and~\ref{multi5876_8}, we have drawn the
$I_{r}$ ratio versus the number of threads along the los, for the
He\,{\sc i} $\lambda 10830$~\AA\, and $D_{3}$ multiplets. For the
calculation of this ratio, we considered the emergent specific
intensity integrated over $D_{z}$
as $${{\bar{I}}(\nu)}\,=\,\displaystyle{\frac{\int_{0}^{D_{z}}\,I(\nu,z)dz}{D_{z}}} \, .$$
The total optical thickness, integrated the same way, is also
displayed on the same plots.

We can thus determine the number of threads required, with such
models, to obtain a $I_{r}$ ratio around 6 for the He\,{\sc i} $D_{3}$
multiplet. In that case, a good value is around 30 threads, which
corresponds, however, to a very large total geometrical width of
$36\,000$ km for the prominence. The corresponding optical thickness
is $\tau(D_{3})\,\sim\,0.2$. For a temperature $T\,=\,17\,000$ K,
corresponding to a prominence to corona transition region (PCTR, see
e.g., Fontenla et al.~\cite{fontenlaetal}, Anzer \&
Heinzel~\cite{anph} and references therein), the number of threads
falls to 15, and $\tau(D_{3})\,\sim\,0.7$.

Even though it is still preliminary, yet we favour multi-thread models
in order to explain a number of properties of observed He\,{\sc i}
spectral lines.

However, we are also aware of the fact that, even in a non-magnetic
field regime, Stokes $I$ is affected by atomic polarization
(R. Casini, private communication). Therefore, a more detailed
comparison with spectropolarimetric data would require a more complex
treatment of the statistical equilibrium.


\section{Conclusion}

We have performed here several tests in order to compare our new 2D
radiative model against previous works. Results for the hydrogen atom
have been compared to Heasley \& Milkey~(\cite{hm83}), Gouttebroze et
al.~(\cite{ghv}) and Paletou~(\cite{fp95}). Results for the helium
atom have been mainly compared to Labrosse \&
Gouttebroze~(\cite{lg01}, \cite{lg04}). An excellent agreement is
reached using our 2D code in the limit of a 1D geometry, with similar
atomic models and incident radiation.

We have shown that, taking into account the atomic fine structure for
the $2^{3}P$, $3^{3}P$ and $3^{3}D$ levels of He\,{\sc i} in our 2D
radiative transfer code, allows to compute emergent line profiles
which can be directly compared to high-spectral resolution
observations. However, using a classical model of isothermal,
isobaric, homogeneous and static slabs leads to ratios between the
subcomponents of He\,{\sc i} $\lambda 10830$~\AA\, and $D_{3}$
multiplets which do not always support observed values, except for
high temperature and high pressure conditions, which are not typical
of quiescent prominences (Engvold et al.~\cite{engvold}).

This led us to undertake a preliminary investigation of 2D
multi-thread modelling. It has indeed given us good hints about how it
should be possible to reproduce observed characteristics of the
$D_{3}$ and $\lambda 10830$~\AA\, multiplets of He\,{\sc i}, such as
the ratio between the intensity of their two subcomponents.

Our 2D radiative model will now be more intensively used, taking into
account smaller threads, eventually beyond spatial resolution (see
Heinzel~\cite{ph07} for a recent review) and the radiative interaction
between them (e.g., Heinzel~\cite{ph89}).  We should also include a
PCTR for each individual thread in our model, since its influence on
hydrogen lines have already been evaluated.

We finally anticipate that our results will be very valuable for the
analysis of data coming from a variety of spaceborne (e.g., SoHO,
Hinode) and groud-based telescopes such as TH\'eMIS.

\begin{acknowledgements}
  We are grateful to Drs. Juan Fontenla (LASP, U. of Colorado),
  Nicolas Labrosse (U. Glasgow) and Petr Heinzel (Ond\v rejov) for
  fruitful discussions, suggestions and comments they provided to us
  during the course of this study. TH\'eMIS is operated on the Island
  of Tenerife by CNRS-CNR in the Spanish \emph{Observatorio del Teide}
  of the \emph{Instituto de Astrof\'{\i}sica de Canarias}.
\end{acknowledgements}

\end{document}